\documentclass{elsart1p}
\usepackage{graphicx}
\usepackage{amssymb}
\begin{document}
\begin{frontmatter}
% notes:
% use the corauthref command within \author for corresponding author
% footnotes;
\title{Hot and dense matter in quark-hadron models}

\author{S. Schramm and J. Steinheimer}

\address{Frankfurt Institute for Advanced Studies, Goethe-University Frankfurt, Germany}

\begin{abstract}
We present a general approach to incorporate hadronic as well as quark degrees of freedom in a unified approach. 
This approach implements the correct degrees of freedom at high as well as low temperatures and densities. 
An effective Polyakov loop field serves as the order parameter for deconfinement. We employ a well-tested hadronic flavor-SU(3) model based on a chirally symmetric formulation that reproduces properties of ground state nuclear matter and yields good descriptions of nuclei and hypernuclei. 
Excluded volume effects simulating the finite size of the hadrons drive the transition to quarks at high temperatures and densities. 
We study the phase structure of the model and the transition to the quark gluon plasma and compare results to lattice gauge calculations. 
\end{abstract}

\end{frontmatter}

% main text
\section{Introduction}
\label{}
A central task of ultra-relativistic heavy-ion collisions is to investigate phase transitions of the strongly interacting matter that is created in the fireball of the collision. Experimentally the investigated parameters of temperature and chemical potential at which possible phase transitions are probed, can be varied by studying heavy-ion collisions at different beam energies, from LHC and RHIC energies with very low net baryon density to lower energies that will be especially investigated at the FAIR facility at GSI which will probe higher densities or chemical potentials, respectively.
The most important phase transitions occurring in hot and dense matter are the restoration of chiral symmetry and the deconfinement transition.
In order to model these transitions theoretically the main problem arises from the very different degrees of freedom in the limit of low and high temperature/density.
At high excitation energy the system is described in terms of quarks and gluons, whereas at low excitation (or as limiting cases, in the vacuum or in the nuclear matter ground state) the effective degrees of freedom are hadrons. In order to describe non-trivial phase transition behavior it is therefore necessary to develop a model that contains both sets of degrees of freedom with the correct asymptotics. In this article we present a model of that kind, the Hadron-Quark-Model (HQM). We investigate the phase transition and present comparisons to lattice gauge calculations at vanishing chemical potential.

\section{The HQM model}

The underlying hadronic SU(3) model has the following structure (see \cite{Papazoglou:1998vr} for details).
In mean field approximation one has
\begin{equation}
L = L_{kin}+L_{int}+L_{meson}.
\end{equation}
$L_{int}$ contains the interactions of the baryons and meson fields:
\begin{equation}
L_{int}=-\sum_i \bar{\psi_i}[\gamma_0(g_{i\omega}\omega+g_{i\phi}\phi)+m_i^*]\psi_i.
\label{formel1}
\end{equation}
The effective baryon masses $m_i^*$ read
\begin{equation}
m_i^* = g_{i\sigma} \sigma + g_{i\zeta} \zeta + \delta m_i 
\label{formel1}
\end{equation}
including couplings to the scalar field plus a small explicit mass term.
$L_{mesons}$ includes the mesonic self-interactions, which in the case of the scalar fields generate the
spontaneous chiral symmetry breaking, as well as self-interactions of vector fields and an explicitly chiral symmetry breaking term: 
\begin{eqnarray}
L_{meson}&=&-\frac{1}{2}(m_\omega^2 \omega^2+m_\phi^2\phi^2)-g_4\left(\omega^4+\frac{\phi^4}{4}+3\omega^2\phi^2+\frac{4\omega^3\phi}{\sqrt{2}}+\frac{2\omega\phi^3}{\sqrt{2}}\right)\nonumber\\
&+&\frac{1}{2}k_0(\sigma^2+\zeta^2)-k_1(\sigma^2+\zeta^2)^2-k_2\left(\frac{\sigma^4}{2}+\zeta^4\right)-k_3\sigma^2\zeta\nonumber\\
&+& m_\pi^2 f_\pi\sigma+\left(\sqrt{2}m_k^ 2f_k-\frac{1}{\sqrt{2}}m_\pi^ 2 f_\pi\right)\zeta~\nonumber\\
&+& \chi^4-\chi_0^4 + \ln\frac{\chi^4}{\chi_0^4} -k_4\ \frac{\chi^4}{\chi_0^4} \ln{\frac{\sigma^2\zeta}{\sigma_0^2\zeta_0}} ~.
\label{formel2}
\end{eqnarray}
The fields $\sigma$ and $\zeta$ denote to the non-strange and strange scalar quark condensates, and $\omega, \phi$ are the corresponding
vector fields. The dilaton field $\chi$ represents the gluon condensate in the system.

In addition the model contains quark degrees of freedom that couple linearly to the mean fields together with a Polyakov loop $\Phi$ field that serves as the order parameter for deconfinement in the spirit of the PNJL model for quarks. We adopt a standard choice of an effective potential for the Polyakov loop \cite {Ratti:2005jh}:
\begin{equation}
	U = -\frac12 a(T)\Phi\Phi^*
	 + b(T)ln[1-6\Phi\Phi^*+4(\Phi^3\Phi^{*3})-3(\Phi\Phi^*)^2]
\end{equation}
 with $a(T)=a_0 T^4+a_1 T_0 T^3+a_2 T_0^2 T^2$, $b(T)=b_3 T_0^3 T$ where the constants are fitted to reproduce quenched lattice results.

Thus effectively we couple a hadronic model with a PNJL model for quarks . In order to naturally suppress hadrons at high densities and temperatures we take into account excluded volume effects in a thermodynamically consistent manner as described in \cite{JSJPHYSG,Rischke:1991ke}. For an alternative way to formulate the HQM model and to suppress hadrons at high temperatures see \cite{Dexheimer:2009hi}.

\section{Results}

We solve the model equations by minimizing the grand canonical potential for given temperature and baryochemical potential.
Figure 1 shows the resulting value of the particle plus antiparticle densities as function of temperature for the case of vanishing chemical potential.
The critical temperature, defined as the maximum of the derivative of the scalar field, has a value of $T_c = 183 $MeV. 
One can see that quarks start to dominate the system at around $T_c$. However, a quite broad range of temperatures can be observed, where the matter
consists of a mixture of quarks and hadronic degrees of freedom. This is a rather natural outcome given the smooth cross-over transition resulting
from this model calculation as well as lattice simulations \cite{Borsanyi:2010cj,Bazavov:2009zn}.
Thermodynamical quantities like the interaction measure also compare well to lattice results \cite{JSJPHYSG}.
\begin{figure}
\begin{center}
\leavevmode
\includegraphics[width=9cm]{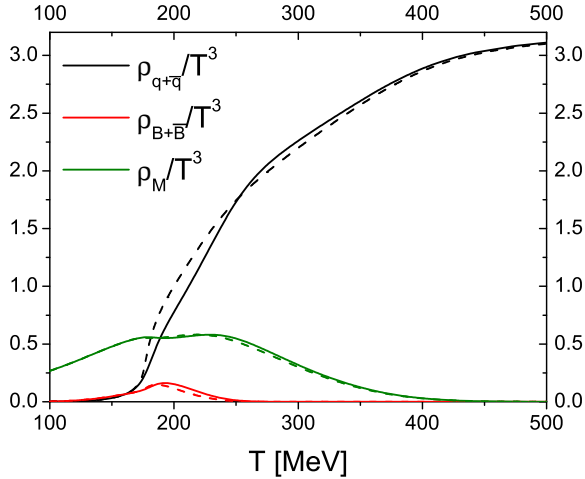}
\end{center}
\caption{\label{dens}Total particle number densities for the different particle species divided by $T^3$ as a function of $T$ at $\mu_B = 0$  \cite{JSJPHYSG}. The black line shows the total number of quarks+antiquarks per volume while the green (dotted) line refers to the total meson density and the red (dashed) line to the number density of hadronic baryons+antibaryons.}
\end{figure}

Studying the phase diagram as function not only of $T$ but also of the chemical potential yields results shown in Fig. 2.
For the parameters used in this study the corresponding nuclear ground state does not contain quarks but is purely governed by the hadronic chiral Lagrangian.
The phase transition is a cross over for all values (this might change, however, if one includes more hadronic resonances).
One can see the transition to chiral restoration as well as a first order liquid-gas phase transition which continues after
a temperature of $T = 16$~MeV as a cross over and which joins the other chiral transition at lower chemical potential and
higher temperatures. Also shown in the figure is the phase transition line to deconfinement, which stays at higher temperatures
for large chemical potential.

Various studies of simulations of heavy-ion collisions implementing the equation of state of the HQM model have been performed 
\cite{JSJPHYSG,dileptons} that show the importance of the quark phase, for instance in the case of dilepton production. More calculations
in this direction are in progress.

%\begin{figure}
%\includegraphics[width=9cm]{figs/fig1a_mu0_sigma_pol.eps}
%\caption{\label{mu0sig}The normalized order parameters for the chiral (red lines), and deconfinement (black line) phase transition as a function of $T$ at $%mu_B = 0$ \cite{JSJPhysG}. Also indicated is the subtracted chiral condensate as defined in the text (orange dash-dotted line). The corresponding lattice data (symbols)  for different lattice actions (asqdat and p4) and lattice spacings $N_{\tau}$ are taken from \cite{Bazavov:2009zn}.
%}
%\end{figure}

\begin{figure}
\begin{center}
\leavevmode
\includegraphics[width=11cm]{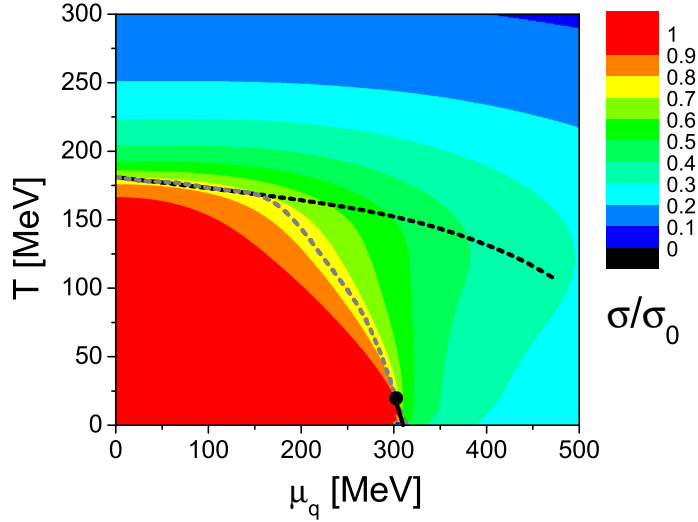}
\end{center}
\caption{\label{tmusiggo}
Contour plot of the chiral condensate normalized to the vacuum value as a function of temperature and quark chemical potential including quark repulsive interactions using a value of the interaction strength of $g_{q \omega} = 3$ \cite{JSJPHYSG}. 
The dashed grey line indicates where the change of the chiral condensate with respect to $T$ and $\mu_q$ has a maximum. 
The dashed black line shows the maximum of change of the Polyakov loop. Here, there is a well-defined nuclear ground state without quarks. The corresponding liquid gas phase transition is shown as the black solid line with a critical endpoint at $T_{cep} \approx 16$~MeV.}
\end{figure}

\end{document}